%% file: elec_funding.tex
\documentclass[12pt]{article}

\input{boilerplate}

\usepackage{todonotes}

\title{\bf Did Private Election Administration Funding Advantage Democrats in 2020?\thanks{For excellent research assistance, the authors thank Jessica Persano. For helpful discussion and comments, the authors thank Avi Acharya, Avi Feller, Anthony Fowler, Justin Grimmer, Christian Grose, Jens Hainmueller, Eitan Hersh, Guido Imbens, Ethan Kaplan, Lynn Vavreck, and Stefan Wager, seminar participants at Dartmouth, UCLA, and Stanford, and attendees of the 2023 Election Science, Administration, and Reform Conference and 2023 Meeting of the Society of Political Methodology. Daniel Thompson also thanks the UCLA Academic Senate's Faculty Research Grant program for generously funding this research.}}

\author{
Apoorva Lal\footnote{Independent Researcher, lal.apoorva@gmail.com, \url{https://apoorvalal.github.io} }
\\[2mm]
Daniel M. Thompson\footnote{Assistant Professor, Department of Political Science. danmckinleythompson@gmail.com, \url{http://www.danmthompson.com}.}
\
}

\IfFileExists{upquote.sty}{\usepackage{upquote}}{}
\begin{document}

\etocdepthtag.toc{mtchapter}
\etocsettagdepth{mtchapter}{section}
\etocsettagdepth{mtappendix}{none}

\maketitle
\thispagestyle{empty}

\begin{abstract}

Private donors contributed more than \$350 million to local election officials to support the administration of the 2020 election. Supporters argue these grants were neutral and necessary to maintain normal election operations during the pandemic, while critics worry these grants mostly went to Democratic strongholds and tilted election outcomes. These concerns have led twenty-four states to restrict private election grants. How much did these grants shape the 2020 presidential election? To answer this question, we collect administrative data on private election administration grants and election outcomes. We then use new advances in synthetic control methods to compare presidential election results and turnout in counties that received grants to counties with identical average presidential election results and turnout before 2020. While counties that favor Democrats were much more likely to apply for a grant, we find that the grants did not have a noticeable effect on the presidential election. Our estimates of the average effect of receiving a grant on Democratic vote share range from 0.02 percentage points to 0.36 percentage points. Our estimates of the average effect of receiving a grant on turnout range from -0.03 percentage points to 0.13 percentage points. Across specifications, our 95\% confidence intervals typically include negative effects, and our confidence intervals from all specifications fail to include effects on Democratic vote share larger than 0.58 percentage points and effects on turnout larger than 0.40 percentage points. We characterize the magnitude of our effects by asking how large they are compared to the margin by which Biden won the 2020 election. In simple bench-marking exercises, we find that the effects of the grants were likely too small to have changed the outcome of the 2020 presidential election.

\end{abstract}

\pagebreak

\pagebreak
\setcounter{page}{1}

\doublespacing

\section{Introduction}\label{sec:intro}

Private donors contributed more than \$350 million to local election officials to support the administration of the presidential election in 2020.\footnote{See \url{https://www.techandciviclife.org/our-work/election-officials/grants/} and \url{https://pollingaccessgrants.org/}} This nearly matches the supplemental funding Congress appropriated to support local election administration in 2020\footnote{\href{https://www.eac.gov/payments-and-grants/CARES}{https://www.eac.gov/payments-and-grants/CARES}} and is a substantial share of the \$2 to \$3 billion spent in a typical national election \citep{mohr2018election}.\footnote{For a review of recent estimates of the cost of elections, see \url{https://electionlab.mit.edu/sites/default/files/2022-05/TheCostofConductingElections-2022.pdf}.} The private donors and their supporters argue this money was necessary to ensure all eligible citizens had a chance to vote amid the disruptions caused by the COVID-19 pandemic.\footnote{\url{https://www.techandciviclife.org/2020covidsupport/}} Many local election officials echo this view with one anonymous official saying they used the funding to ``alleviate choke points and barriers to voting.''\footnote{\url{https://www.techandciviclife.org/election-officials-made-democracy-happen-in-2020/}} Favoring these arguments, Michigan voters approved an amendment to their state's constitution in 2022 that protects the right of local governments to receive private funding for election administration.

Critics of these donations argue that private donors overwhelmingly favored Democratic-leaning counties and municipalities and that these grants have the potential to tilt elections in favor of one party by increasing the participation of some citizens more than others.\footnote{See, e.g., \url{https://thefga.org/briefs/show-me-the-zuckerbucks-outside-money-infiltrated-missouris-2020-election/}} In one complaint filed before the Federal Election Commission, the plaintiff points out that a large share of the funds donated in 2020 went to Democratic-leaning parts of the country and alleges that the donors have a ``hidden motive to increase Joe Biden’s statewide vote.''\footnote{\url{https://eqs.fec.gov/eqsdocsMUR/7854_01.pdf}} These concerns led twenty-four states to adopt laws banning or limiting private donations to local election officials.\footnote{\url{https://www.ncsl.org/elections-and-campaigns/prohibiting-private-funding-of-elections}} How much did private election administration grants tilt the 2020 election?

We address this question by combining county-level administrative data on turnout, presidential voting, election spending, and election administration with records of which county governments received a private donation from the largest private election administration donor in 2020, the Center for Tech and Civic Life (CTCL).
We document that %
counties that support Democrats were more likely to apply for private election funding in 2020, which is consistent with critics' allegations.  Since Democratic-leaning counties were much more likely to apply, and every eligible applicant received a grant, a simple comparison of turnout and presidential vote shares in counties that did or did not receive a grant fails to reveal the grant's impact. We mitigate this bias by comparing grant-funded counties to those without funding but with similar pre-2020 turnout and voting trends using recent advancements in synthetic control methods \citep{arkhangelsky2021synthetic}.

We find that, despite the scale of the CTCL grant program in 2020 and the tendency of the money to go to Democratic-leaning counties, private funding did not noticeably advantage Joe Biden in the 2020 presidential election. We estimate that receiving a grant increased support for Democrats by between 0.02 and 0.36 percentage points and increased turnout by less than 0.13 percentage points. We validate our estimates using alternative machine learning and econometric approaches to estimating the effects, and these approaches produce similar estimates. Since large counties and counties in battleground states have a larger effect on the aggregate election outcome, we also estimate the effects in these counties separately. We find that grants had a similar effect in large counties and counties in battleground states. We also present evidence that the small average effects are not masking large effects for the small number of counties that received relatively large grants.

To characterize the magnitude of the effects, we compare our estimates to the state-by-state margins in the 2020 presidential election. We conduct a simple analysis in which we remove the average effect on turnout and Democratic vote share from every grant-receiving county's vote total. Despite the razor-thin margins in 2020, we find that the turnout and Democratic vote share effects are not large enough to have swung the election to Donald Trump.

Beyond the ongoing policy debate about private election funding, this paper contributes to a growing social scientific literature on the effects of local election administration. Democratic and Republican officials often disagree over how much to spend on elections and how they should be funded \citep{hasen2012voting,mohr2019strategic}. This leads to a conventional wisdom that spending more or less will have substantial effects on the outcomes of elections. Yet, while some changes in state and local election administration can affect participation and alter the composition of the electorate, the turnout effects tend to be modest, and the compositional effects are often hard to predict  \citep{cantoni2021strict,clinton2020polling,gerber2013identifying,gronke2008convenience,kaplan2020early,tomkins2023blocks,thompson2020universal,yoder2021did}. Further, despite large differences between the Democratic and Republican positions on how much money to spend on elections, Democratic local officials do not produce more turnout or higher Democratic vote shares than do Republican local officials \citep{ferrer2023partisan}. This paper advances this literature by evaluating a large increase in resources rather than a single policy change. If local election officials are motivated to increase participation and are well informed about what administrative changes will be most effective, providing them resources should have a larger effect on turnout than any individual policy change. Our findings suggest that there may be less room for increasing turnout with local administrative changes than previously expected.

This paper also contributes to the large literature on the role of money in American politics. A vast literature studies the effects of campaign finance and the influence of special interests \citep[see, e.g.,][]{ansolabehere2003there,levitt1994using}. This project is especially closely related to a small but growing literature on the influence of business leaders in politics \citep{hersh2023political}. Many reforms have sought to limit the influence of money in politics by changing who can spend how much money on which races \citep[see, e.g.,][]{fouirnaies2022campaign,hall2016systemic,kilborn2022public,yorgason2021campaign}. While many public officials and members of the public are concerned that donations to support local election administration offer a new, previously untapped way to influence election results without giving directly to candidates, our results suggest this may not be a substantial risk.

When considering the implications of our results, it is important to note that our paper reports estimates of the average effect of additional election funds at the current margin. If, for example, election officials had much smaller budgets than in 2020, grants may help to maintain the most basic election functions and thereby have substantial effects on participation and election outcomes. More resources may also help local governments maintain better security measures or make voting more convenient. While we cannot measure this directly, many of the recipients said the grants helped them make their local election more secure.\footnote{For example, one election official told the CTCL: ``We are a small community struggling to find ways to handle unfunded mandates, especially during a pandemic. It means a lot to us to ensure our election process is done in all the right ways.'' \url{https://www.techandciviclife.org/grant-update-october/}}

\section{Private Election Administration Grants in 2020}

\subsection{Center for Tech and Civic Life's 2020 Grant Program}\label{sec:grant_details}

In fall 2020, Mark Zuckerberg and Priscilla Chan donated approximately \$350 million to the Center for Tech and Civic Life (CTCL), a Chicago-based nonprofit, to administer a grant program for local election administration. The CTCL invited all local governments responsible for administering elections to apply for funding and gave the funding to every eligible government that applied.\footnote{Unless cited to a different source, we rely on the CTCL's website for details about the grants. \url{https://www.techandciviclife.org/our-work/election-officials/grants/}} The grants were intended to offset election administration expenses incurred from June 2020 until December 2020. The CTCL determined the maximum amount of each grant based on the eligible voting population of each jurisdiction as well as other demographics. According to our calculations, CTCL gave the median grant-receiving county approximately \$0.81 per voting-age resident. The typical local government spends approximately \$8 per eligible citizen on elections \citep{mohr2018election}, making this a roughly 10\% increase in the typical recipient's election funding in a normal year.

According to reports submitted to the CTCL, election officials intended to use the grants in a variety of ways to make election day run smoother, offer alternative ways to vote, and reduce COVID transmission.\footnote{\url{https://www.techandciviclife.org/grant-update-november/}} Officials reported using the money to hire poll workers and other temporary staff, purchase mail balloting equipment and supplies, obtain protective equipment such as masks, and purchase other standard election equipment. One election administrator told the CTCL that ``this unprecedented voter participation simply would have crippled the administration of our elections with devastating effects if we were left with the limited available municipal funds.''\footnote{\url{https://www.techandciviclife.org/election-officials-made-democracy-happen-in-2020/}}

\subsection{Grant and Election Data}\label{sec:data}

\begin{figure}[t]
\label{fig:map}
\caption{\textbf{Geographic Distribution of Grants} }
\centering
\vspace{-10mm}
\includegraphics[width=\textwidth]{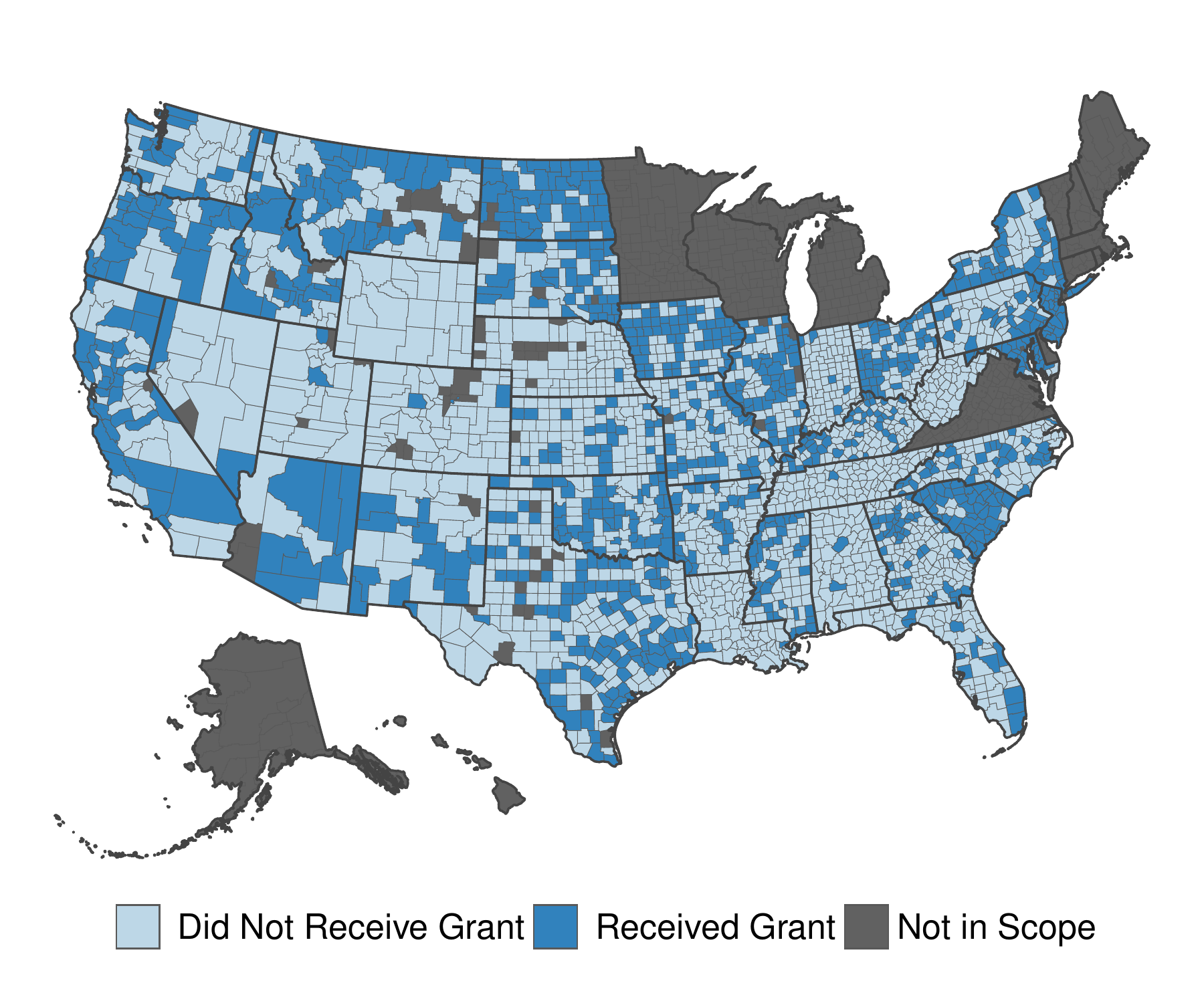}
\label{fig:vs_selection2}

\end{figure}

We draw on grant and election data to study the effect of private election administration grants in the 2020 election. We build our dataset of grant recipients using the CTCL's 2020 tax filing which contains a list of every grant made under this program. We digitized this tax document and extracted all of the recipient names and grant amounts. We also compiled and cleaned county-level presidential election results from 1992 to 2020. These results were collected from secretaries of state and reported in Dave Leip's Atlas of US Presidential Elections.\footnote{Alaska's secretary of state reports election results at the election district level rather than the borough level, which is the equivalent of counties in Alaska and the level at which CTCL made grants in Alaska. We exclude Alaska from our data.}

The official responsible for running elections varies across states and even, occasionally, within states \citep{kimball2006street}. In ten mostly New England and Midwestern states, election administration is largely handled by municipal governments.\footnote{These states are Connecticut, Maine, Massachusetts, Michigan, Minnesota, New Hampshire, Rhode Island, Virginia, Vermont, and Wisconsin.} In these states, we cannot distinguish between counties that did and did not receive grants because there are often many municipalities within a given county. To avoid incorrectly labeling these counties as receiving a grant when only a small portion of the county received a grant, we withhold these states from our analysis. We also withhold five counties with municipal election administrators in states where elections are typically run by county officials.\footnote{We exclude Cook, St. Clair, Vermilion, and Winnebago counties in Illinois and Jackson County in Missouri.} Finally, we exclude 59 counties that either have fewer than 1,000 residents or have changing borders during our analysis period given the challenges associated with estimating turnout when the population estimate in the denominator of turnout will be noisy.

\subsection{Reasoning About the Effects of Election Administration Grants}

Should we expect grants like this to affect turnout and election results? One way to reason about this is to consider each of the changes in election administration that the grant money facilitates and evaluate who the change targets, how much it will increase participation in the targeted group, and how different the targeted and non-targeted groups are in terms of expected partisan voting.\footnote{This section draws heavily from \cite{grimmer2023election} and private conversation with the authors.}

Consider a grant-receiving county that spends the money on additional poll workers---more counties mentioned spending grant money on temporary staff or poll workers than on any other spending category.\footnote{\url{https://www.techandciviclife.org/grant-update-november/}} Staff may make the process of registering or requesting an absentee ballot easier, activities that are necessary for well-run elections but are unlikely to have substantial effects on participation. They may also help keep lines shorter. Suppose the additional staff reduce the average wait time by 10 minutes, a quite large effect. \cite{pettigrew2021downstream} finds that wait times decrease future turnout by approximately 1 percentage point for every hour a person waits. If this effect is linear and applies to people deciding whether to stay or leave based on line length, not just in future years, then reducing wait times by 10 minutes for the average voter increases participation by 0.17 percentage points. Imagine there are two polling places in a county with equal numbers of voters using each in a typical year. The local election official sends the new staff to only one location. Suppose further, as an extreme example, that 75\% of voters in the precinct that received the extra staff typically vote for Democrats while only 25\% of voters in the other precinct vote for Democrats. In this extreme example, overall turnout increases by 0.085 percentage points and Democratic vote share increases by an even smaller 0.021 percentage points. Similar exercises using a similar approach for other common interventions---like adding polling places \citep{clinton2020polling, tomkins2023blocks, yoder2018polling}, adding ballot drop boxes, and expanding early voting hours \citep{gronke2008convenience,kaplan2020early}---lead to the same conclusions. A large share of counties also reported spending money in ways that should have even smaller effects on participation and partisan balance like purchasing personal protective equipment for poll workers or purchasing new election equipment.

Even after combining many small effects, this exercise leads us to expect that the effect of this grant money on turnout and Democratic vote share is quite small if there is any detectable effect at all.

\section{Democratic-Leaning Counties Were More Likely to Apply for and Receive a Grant}\label{sec:selection}

One of the key concerns among critics of the private election administration grants is that Democratic-leaning counties were more likely to receive them. The CTCL gave a grant to every eligible county and municipality that applied,\footnote{The FEC held that this fact was uncontested in MUR 7854. \url{https://www.fec.gov/files/legal/murs/7854/7854_25.pdf}} but every county and municipality did not apply for the funding. If Democratic counties are more likely to apply and receive the money, and the money leads to higher turnout, the money can advantage Democrats when the state adds up the county totals.

\begin{figure}[t]
\caption{\textbf{Democratic-Leaning Counties More Likely to Apply for and Receive Election Assistance Grants.} Each dot represents the average of 144 counties binned based on two-party Democratic presidential vote share from 2016. The regression line is fit to the underlying county-level data.\label{fig:selection_demvs}}
\centering
\includegraphics[width=0.7\textwidth]{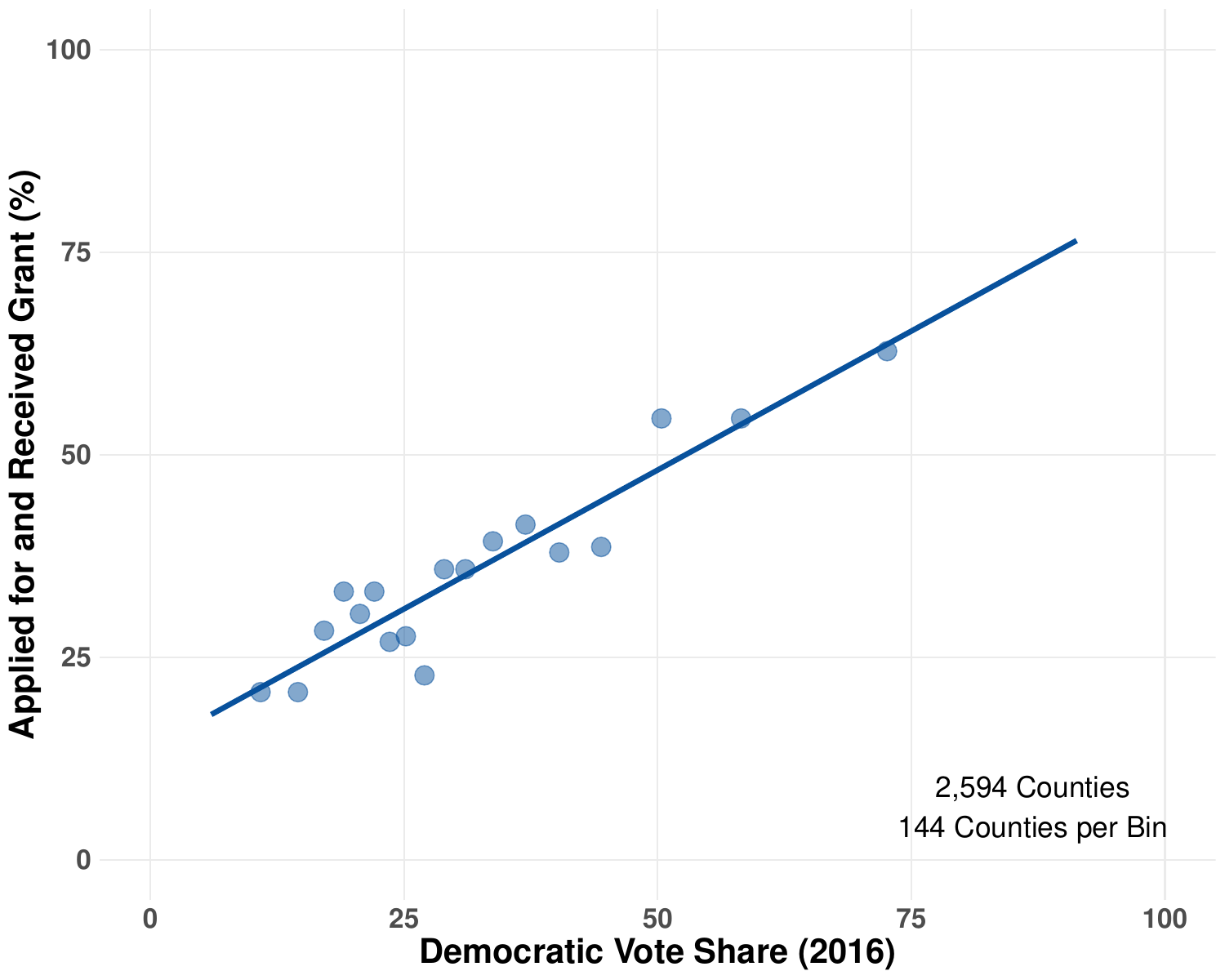}
\end{figure}

Figure \ref{fig:selection_demvs} captures how much more likely Democratic-leaning counties were to apply for a grant than Republican-leaning counties. We use Democratic vote share in the 2016 presidential election as a measure of partisan lean. We find that the CTCL gave grants to about 30\% of counties where Donald Trump received approximately 75\% of the two-party presidential vote in 2016. Meanwhile, the CTCL gave to 65\% of counties where Hilary Clinton received 75\% of the two-party vote in 2016.

\subsection{Why Did Democratic-Leaning Counties Apply at Higher Rates?}

Democratic-leaning counties differ from Republican-leaning counties in a large number of ways that could plausibly make them more likely to apply for a private election grant: they have larger populations, they are more densely populated, they are more racially and ethnically diverse, and they were more exposed to COVID-19 before the grants were announced on average. Might these characteristics account for the tendency of Democratic counties to apply for the private election grants at higher rates than Republican counties?

\input{figtab/selection_table.tex}

In Table \ref{tab:selection}, we document some simple descriptive patterns to help us understand how plausible are different explanations for the Democratic-Republican gap in private election grant applications. Each column presents coefficient estimates from a linear regression of grant receipt on 2016 Democratic presidential vote share and, in some cases, additional factors that might correlate with Democratic vote share and applying for a private election grant. We do not intend to estimate the causal effect of changing the share of Democrats in a county on the probability of applying for a grant. Instead, we use these regressions to evaluate plausible explanations for the Democratic-Republican gap in grant application rates.

As a point of reference, column 1 of Table \ref{tab:selection} reports the coefficient on 2016 Democratic vote share from a simple bivariate regression. The coefficient implies that the probability of applying for and receiving a grant is 6.8 percentage points higher on average in a county with a 10-percentage-point higher Democratic vote share.

Some states were more favorable to private election grants while others actively discouraged counties from applying. For example, 23 states applied for private election grants from a similar grant program for state-level election officials.\footnote{For a discussion of these grants, see \url{https://electioninnovation.org/research/ceir-2020-voter-education-grant-program/}.} Some state officials also reached out to local officials encouraging them to apply while others threatened to sue local officials that applied.\footnote{See the case of Louisiana in which the Secretary of State encouraged applications while the Attorney General threatened to sue parishes that applied. \url{https://www.theadvocate.com/baton_rouge/news/politics/elections/mark-zuckerberg-funded-free-election-grants-draw-ire-of-jeff-landry-who-files-suit/article_e59425a0-08a2-11eb-9757-cba83bb12048.html}} In column 2 of Table \ref{tab:selection}, we include state fixed effects and find that the coefficient on 2016 Democratic vote share is approximately 19\% lower after accounting for the tendency of Democratic counties to be in states where counties applied for the CTCL grant at a higher rate.

Among counties in the same state, Democratic counties tend to have more people, be more urban and racially diverse, be members of a national organization that informed counties about the grants, and have had worse experiences with COVID-19 in the spring and summer of 2020. We evaluate whether these differences between Democratic and Republican counties can account for the pattern of Democratic counties applying for the private election grants at higher rates. In column 3 of Table \ref{tab:selection}, we find that counties were 3.1 percentage points more likely to apply than other counties in the same state with the same populations and median incomes that voted 10 percentage points more for Republicans in 2016. In column 4, we add to the regression in column 3, adjusting for the non-Hispanic-white share of the population and whether the county is urban or rural. After accounting for population, income, and the state a county is in, urban counties with a smaller share of non-Hispanic white residents vote more for Democrats and applied for the election administration grant at a lower rate. Accordingly, adjusting for the fact that Democratic counties are more urban and diverse, we find that the coefficient on 2016 Democratic vote share increases to 0.44. In columns 5 and 6 we make further adjustments to account for COVID-19 deaths and social distancing preferences prior to the 2020 election as well as membership in the National Association of Counties (NACo) which notified members of the grants. Accounting for these factors does not substantially change the coefficient on 2016 Democratic vote share.

In Section \ref{sec:selection_covid} in the online appendix, we present an additional analysis in which we compare the relationship between 2016 Democratic vote share and applying for a grant in states that dramatically expanded mail balloting in 2020 compared to states that severely restricted mail balloting or already mailed every registrant a ballot. The estimates are noisy and sensitive to which covariates we include in the regression, but the point estimates are consistent with some election officials in Democratic counties anticipating an increase in their costs associated with a transition to mail balloting, and applying for a grant to cover those costs.

\FloatBarrier

\section{Grants Did Not Substantially Increase Turnout or Democratic Vote Share}

In this section, we detail our finding that the private election administration grants did not substantially increase turnout or Democratic vote share, and we explain how we estimate the effects of the grants. This section has five parts: First, we describe our estimation strategy and why it is appropriate for this setting. Second, we present graphical evidence that grants did not substantially increase turnout or Democratic vote share. Third, we report estimates of the effect of grants on turnout and Democratic vote share. Fourth, we document our independent analysis of grants to municipalities in Wisconsin producing similar results as our main estimates. Fifth, we discuss alternative estimation strategies and document how all of these strategies yield similar results. Sixth, we present evidence that the effects are similar in more and less competitive states and more and less populous counties. Finally, we document that the effects are less positive in counties that received larger grants, suggesting that funding is not substantially affecting turnout or the composition of the electorate at the current margin.

\subsection{Estimating the Effect of Private Grants in the 2020 Election}\label{sec:methods}

\begin{figure}[t]
\caption{\textbf{Democratic Vote Share Was Declining Slower and Turnout Was Increasing Faster in Grant-Receiving Counties Long Before 2020.}}\label{fig:trends}
\begin{subfigure}[b]{0.52\textwidth}
\includegraphics[width=\textwidth]{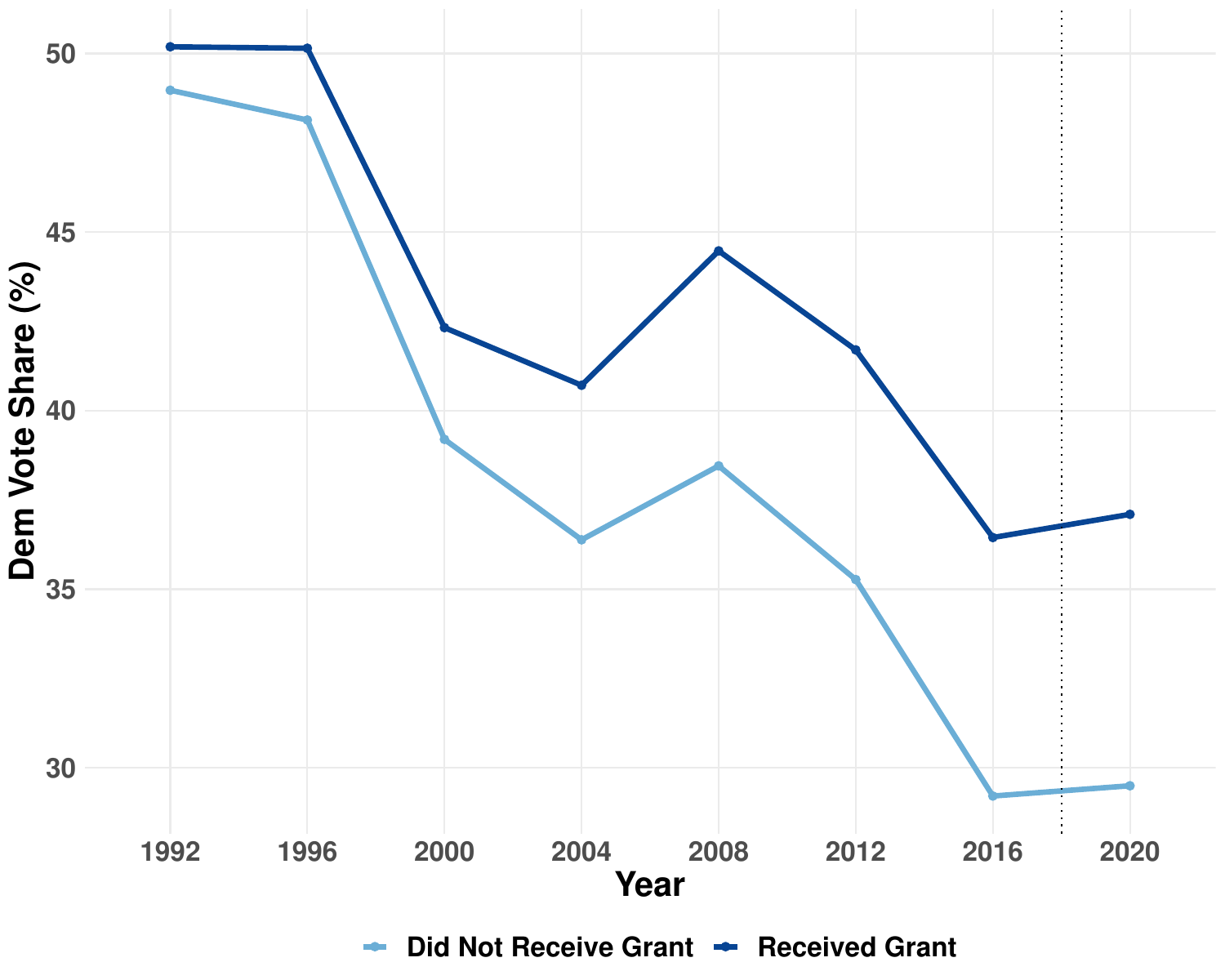}
\label{fig:vs_selection}
\end{subfigure}
\hfill
\begin{subfigure}[b]{0.52\textwidth}
\includegraphics[width=\textwidth]{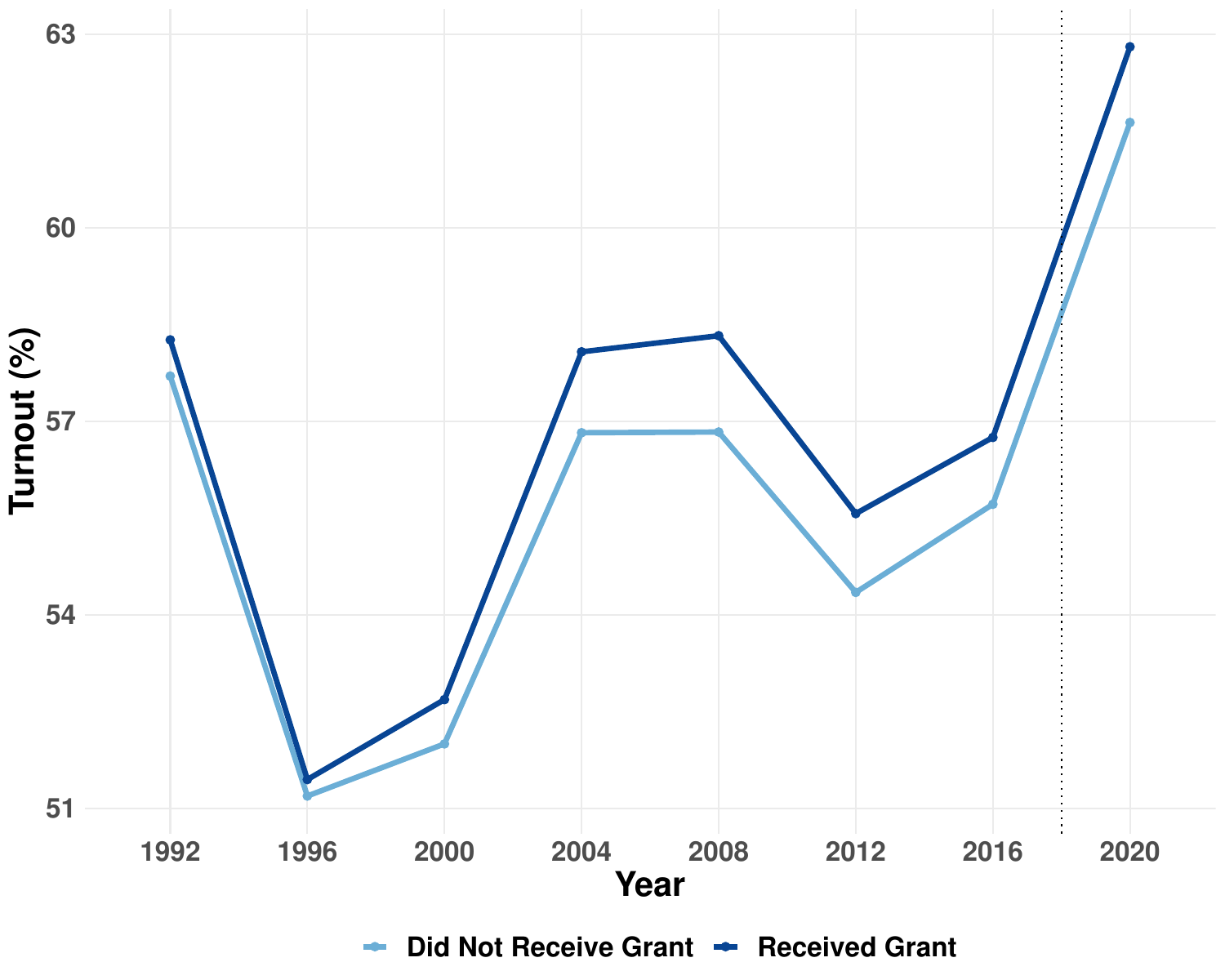}
\label{fig:tur_selection}
\end{subfigure}
\end{figure}

Our goal in this section is to estimate the average effect CTCL grants had on turnout and Democratic vote share. As we document in Figure \ref{fig:selection_demvs}, grant-receiving counties favored Democrats in 2016, four years before the grants were made. Given the tendency of counties to continue voting for the same party from one election to the next, we would expect grant-receiving counties to favor Democrats in the 2020 election more than counties that did not receive grants even if the grants had no effect on Democratic vote share or turnout. This type of selection can be accounted for using a difference-in-differences strategy, where we compare the difference between 2020 and pre-2020 turnout and vote share for Democrats in treated counties with the analogous difference in untreated counties. This approach would yield valid causal estimates under the assumption that turnout and Democratic vote share would have increased by the same amount in 2020 in treatment and control counties in the absence of the grants. Since we have measures of county-level turnout and Democratic vote share for many elections prior to 2020, we evaluate the plausibility of this assumption in Figure \ref{fig:trends}. We find that Democratic vote share is decreasing slower in treated counties than in control counties. We also find that turnout is increasing faster in treated counties. These differences in treated and control trajectories suggest that simple difference-in-differences estimates will dramatically overstate the effect of grants on Democratic vote share and turnout.

To address the shortcomings with the standard difference-in-differences design in our setting, we follow \cite{arkhangelsky2021synthetic} in reweighting our difference-in-differences regressions with weights ensure the treatment and control units are on similar trajectories prior to the treatment and make the pre-treatment period as similar as possible to the post-treatment.\footnote{This approach, which they call synthetic difference-in-differences, builds on the synthetic control method and other related approaches \citep{athey2021matrix, abadie2010synthetic, ben2021augmented, xu2017generalized}} This approach has three steps: 1) compute county weights that make the trend in the control units approximately equal to the trend in the treated units, 2) compute election weights that make the pre-treatment period as similar as possible to the post-treatment period among control units, and 3) estimate a reweighted difference-in-differences regression weighting by the product of the county and election weights.

Formally, we compute county weights $\omega_i$ that minimize the expression

$$ \text{argmin}_{\omega_0 \in \mathbb{R}_+, \omega \in \Omega} \sum_{t=1}^{T_{pre}} (\omega_0 + \sum_{i=1}^{N_{co}} \omega_iY_{it} - \frac{1}{N_{tr}} \sum_{i=N_{co}+1}^{N} Y_{it})^2 + \zeta^2 T_{pre}||\omega||_2^2 $$

where $Y_{it}$ is the outcome in county $i$ and election $t$, $T_{pre}$ is the number of pre-treatment periods, $N$ is the number of control and treated units, $N_{co}$ is the number of control units, $\Omega$ is the set containing all $\omega$ in which all $\omega_i$ fall between between 0 and 1 inclusive and $\omega$ sums to one, and $\zeta$ is a regularization parameter proposed in \cite{arkhangelsky2021synthetic}. Since $\omega_0$ is not regularized, $\omega_0$ represents the average pre-treatment difference between the treated and control units and means that the county weights produce a weighted control mean that follows the same trajectory as the treatment mean but may not be at the same level.

We then compute time weights $\lambda_t$ that minimize an expression nearly identical to the expression for the unit weights:

$$ \text{argmin}_{\lambda_0 \in \mathbb{R}_+, \lambda \in \Lambda}
\sum_{i=1}^{N_{co}} (\lambda_0 + \sum_{t=1}^{T_{pre}} \lambda_tY_{it}
- \frac{1}{T_{post}} \sum_{t=T_{pre}+1}^{T} Y_{it})^2 $$

We use the product of these weights as the weights in a weighted difference-in-differences least squares regression

$$ \text{argmin}_{\tau, \alpha, \beta} \sum_{i=1}^N \sum_{t=1}^T (Y_{it} - \alpha_i - \beta_t -W_{it}\tau)^2 \hat{\omega_i}\hat{\lambda}_t $$

where $W_{it}$ is an indicator for the treatment, $\alpha_i$ is a county fixed effect, $\beta_t$ is an election fixed effect, and $\tau$ is the treatment effect estimate.
As with the classical synthetic control method \citep{abadie2010synthetic} and generalized synthetic control method \citep{xu2017generalized}, the weighted regression coefficient $\hat{\tau}$ from the synthetic difference in differences method yields consistent estimates for the average treatment effect on the treated (ATT) under a low-rank approximation for the untreated potential outcome $Y^0$, which requires that election outcomes in the absence of the grants can be approximated by county intercepts, election year shocks, and low-rank time varying slopes at the county level. This design assumption is strictly weaker than the parallel trends assumption required by difference in difference methods, which we see is implausible from figure \ref{fig:trends}.

To validate the weighted difference-in-differences estimator, we study the 2016 presidential election as a placebo case. We do this by deleting 2020 from our data and pretending that the grants were handed out in 2016. We then rerun the weighted difference-in-difference procedure to produce estimates of the placebo effect. Since the grants were not handed out until 2020, an unbiased estimator will find placebo effects that are close to zero. We present these estimates in Section \ref{tab:placebo2016} in the online appendix. Consistent with the goal of the estimator, we confirm that the weighted difference-in-differences approach fails to estimate a statistically significant placebo effect.

\subsection{Graphical Evidence That Grants Had Minimal Effect on Democratic Vote Share and Turnout}\label{sec:graphical}

\begin{figure}
\caption{\textbf{Trends in Democratic vote share and turnout in treated and synth difference-in-differences counterfactual over time}}\label{fig:sdid}
\begin{subfigure}[b]{0.52\textwidth}
\includegraphics[width=\textwidth]{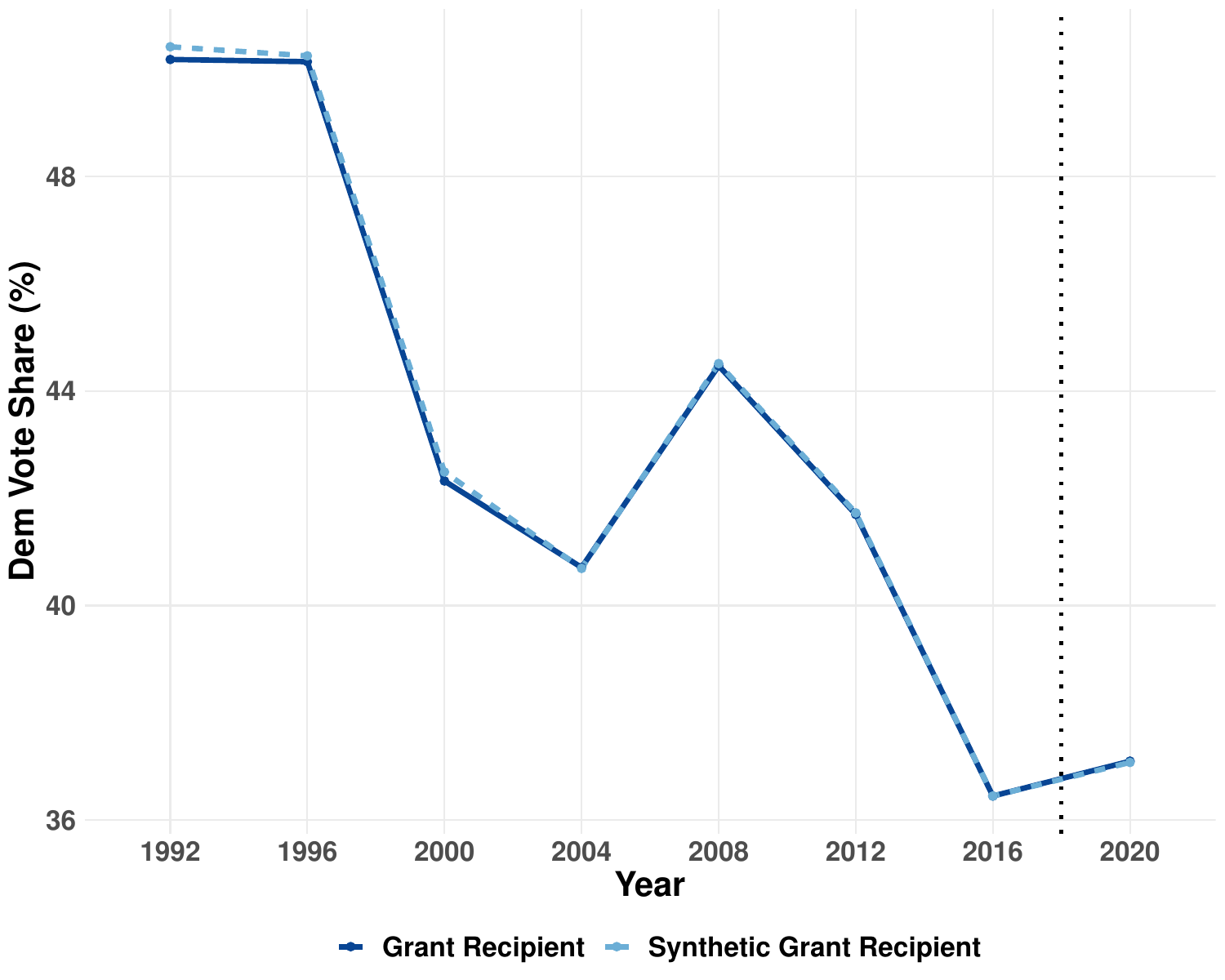}
\caption{Trends in Democratic Vote Share}
\end{subfigure}
\hfill
\begin{subfigure}[b]{0.52\textwidth}
\includegraphics[width=\textwidth]{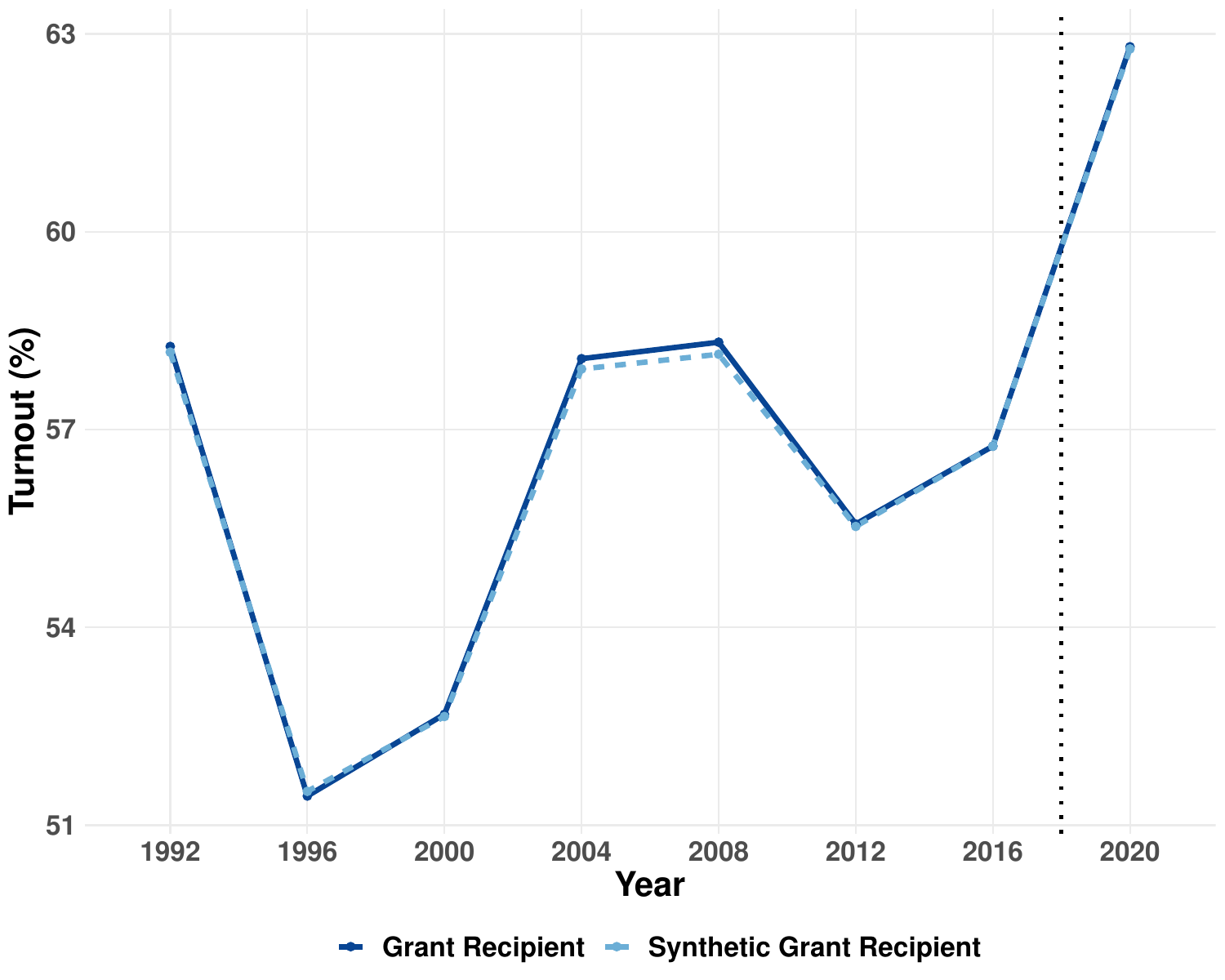}
\caption{Trends in Turnout}
\end{subfigure}
\end{figure}

First, we present graphical evidence that CTCL grants did not substantially increase turnout or vote share for Democrats. Figure \ref{fig:sdid} compares Democratic vote share and turnout for the average grant recipient over time to the counterfactual implied by synthetic difference-in-differences. Across both the Democratic vote share panel on the left and the turnout panel on the right, we see the synthetic difference-in-differences procedure produces a counterfactual that almost perfectly matches the average trajectory for grant recipients. The exceptions to this perfect match are in 1992 where counterfactual Democratic vote share is slightly higher than observed Democratic vote share and 2004 and 2008 where counterfactual turnout is slightly lower than observed turnout. In all three cases, the differences are small with gaps of less than 0.25 percentage points.

Figure \ref{fig:sdid} also makes clear that any average effect of the grants is so small as to not be visible in the Democratic vote share or turnout plots. If there were a visible effect in either plot, it would appear as a difference between the grant recipient and counterfactual lines in 2020. Instead, these lines continue to perfectly overlap in 2020 just like they did prior to 2020, implying that the average grant did not substantially advantage either party or noticeably increase turnout.

\subsection{Estimates of Effect of Grants on Democratic Vote Share and Turnout}\label{sec:effect}

\input{figtab/main_effects_table.tex}

Next, we present our estimates of the effect of receiving a grant on turnout and Democratic vote share. Table \ref{tab:main} reports our estimates using a variety of estimation approaches. Column 1 is the simple difference-in-differences regression estimate of the effect of grant receipt on Democratic vote share. As we establish in Section \ref{sec:methods}, this is a dramatic overestimate of the effect because, even before 2020, Democrats were increasingly performing better in counties that received a grant than in counties that did not. In column 2, we present estimates from difference-in-differences regressions including time weights but not including county weights. The 2020 election result is much more similar to the 2016 result than any other  previous period in our data, so the weight selection procedure places all of the mass on 2016 in the pre-treatment period. This means that column 2 is equivalent to a two-group, two-period difference-in-differences design using only 2016 and 2020. With this estimator, we find that grants increase Democratic vote share by 0.36 percentage points. In column 3, we present our estimate from a difference-in-differences regression with county weights but no time weights. We find that, compared to counties that did not receive a grant but were on a similar average presidential voting trajectory prior to 2020, and after accounting for remaining pre-2020 average differences in Democratic vote share, grant recipients had only a 0.1-percentage-point higher Democratic vote share in 2020. Column 4 presents our difference-in-differences estimate using both time and county weights. Using this specification, we estimate that grants caused an average increase in Democratic vote share of 0.02 percentage points. In summary, once we compare grant recipients to more similar counties, we find that these grants did not substantially increase Democratic vote share in receiving counties.

In columns 5 through 8, we present estimates of the effect of a grant on turnout using the same estimation strategies as in columns 1 through 4. Grant recipients and non-recipients are on more similar turnout trajectories than Democratic vote trajectories prior to 2020. Accordingly, the four estimation strategies produce more similar results. In column 5, we present a likely upwardly biased 0.24-percentage-point difference-in-differences estimate of the effect of the grants on turnout. In columns 6 through 8 we find that, when we use time weights, county weights, or both, our estimates range from -0.03 percentage points to 0.13 percentage points.

Focusing on our preferred specification in columns 4 and 8 where we use both time and county weights in a difference-in-differences regression, we find that the grants did not substantially increase Democratic vote share or turnout. Our estimates are highly precise on both outcomes. The standard error we report in column 4 is 0.11 percent, meaning that we could reject a hypothetical effect of the grants on Democratic vote share that is greater than 0.25 percentage points. Similarly, the standard error we report in column 8 is 0.14 meaning that we could reject a hypothetical effect of 0.30 percentage points.

\FloatBarrier

\subsection{Estimated Effects Similar or Less Favorable to Democrats in Wisconsin}\label{sec:wisc}

In this section, we supplement our main finding with new municipality-level election data from Wisconsin. We find that the grants did not substantially increase Democratic vote share and or turnout.

As we discuss in Section \ref{sec:methods}, our main analyses study places where elections are administered at the county level because most parts of the country administer elections at that level, presidential election data is widely available at the county level, the population denominator is noisily measured in the many small municipalities, and changing municipal boundaries add potentially systematic noise to the election results data. Wisconsin is one of the states we leave out of this analysis because elections are primarily conducted at the municipal level \citep{herron2023allegations}. To validate our main findings, we constructed a separate dataset of municipal-level election results and grant receipt in Wisconsin. We provide more details on this dataset in Section \ref{sec:wisc_appendix} in the online appendix.

Using the same estimation strategies we use in Table \ref{tab:main}, we report estimates of the effect of private election administration grants in Wisconsin in Table \ref{tab:wi} in the online appendix. We estimate very similar effects of the grants on Democratic vote share in Wisconsin as in our nationwide county-level analysis. Our preferred synthetic difference-in-differences estimator with county and year weights in column 4 estimates an effect of 0.11 percentage point. Across all three of our synthetic difference-in-differences estimates, our point estimates range from 0.11 to 0.31 percentage points, and we cannot reject the null hypothesis that the grants had no effect on Democratic vote share. The upper bound of the 95\% confidence interval on our preferred estimate is an effect of 0.48 percentage points.

In columns 6 through 8 of Table \ref{tab:wi}, we report our estimates of the effect of grants on turnout in Wisconsin. Contrary to the expectation that grants increased turnout in Democratic strongholds and thereby advantaged Democrats, we find that municipalities that received grants had a modest but statistically significant drop in turnout of approximately 0.7 percentage points. We interpret this result as evidence against grants improving turnout in Democratic strongholds rather than strong evidence that the grants caused lower turnout in grant-receiving Wisconsin municipalities.

\subsection{Estimated Effects Not Sensitive to Estimation Strategy}\label{sec:alt_est}

We examine the robustness of our main estimates in Table \ref{tab:main} using five alternative estimation strategies. Across all five approaches, we find similar, substantively small effects of the grants on turnout and Democratic vote share. Our estimates of the average effect on Democratic vote share, reported in Table \ref{tab:alt_est_dem} in the online appendix, range from 0.15 percentage points to 0.40 percentage points. Our estimates of the average effect on turnout, reported in Table \ref{tab:alt_est_turnout} also in the online appendix, range from -0.02 percentage points to 0.47 percentage points. These estimates come from five different strategies: 1) weighted regressions like synthetic difference-in-differences but without county-specific intercepts so the weights attempt to balance treatment and control outcomes on levels rather than trends, 2) regularized synthetic control \citep{doudchenko2016balancing}, 3) weighting the control units such that pre-2020 control-group outcome means exactly match treatment means while deviating as little as possible from uniform weights \citep{hainmueller2012entropy}, 4) predicting the outcome and treatment propensity using random forests and using these estimates for augmented inverse propensity weighting \citep{athey2019grf}, and 5) predicting the outcome and treatment propensity using ensemble learners---pooling generalized additive models, boosting, regression trees, splines, and elastic nets---and using these predictions for augmented inverse propensity weighting.

\subsection{Estimated Effect Similar in Battleground States}\label{sec:battleground}

One concern is that, while the grants had a small effect on average, they may have had a larger effect in the closest states. We evaluate this claim by estimating the average effect of the grants on Democratic vote share and turnout in three sets of the most competitive states---states decided by less than 5 percentage points, states that the Cook Political Report identified as battleground states prior to election day, and Wisconsin alone.

In Table \ref{tab:battleground} in the online appendix, we document that, even in the most competitive states, the effect of the grants was small. In column 1, we find that the effect of grants on Democratic vote share in close states was 0.54 percentage points. While this point estimate is larger than the effect we estimate using the full sample, the smaller sample size also means the subgroup estimates are substantially noisier, and we are unable to reject the hypothesis that the grants had no effect. When we extend our analysis to the battleground states according to the Cook Political Report, our estimate is more precise, and the estimated effect on Democratic vote share is nearly identical to our estimates from Table \ref{tab:main} using all counties.

In columns 3 and 4, we find that the effect of the grants on turnout in states decided by less than 5 percentage points and those the Cook Political Report labeled as battlegrounds was approximately the same size as we estimate using the full sample. In both cases, we cannot reject the hypothesis that the grants failed to increase turnout. We can also rule out positive effects on turnout of greater than 0.75 percentage points in the closest states.

Returning to the evidence we presented in Section \ref{sec:wisc}, when we study the battleground state of Wisconsin independently, we find roughly the same pattern of small average effects, no average effects, or even negative average effects of the grants on Democratic vote share and turnout.

\subsection{Estimated Effect Similar in Populous Counties}\label{sec:pop}

Private election funding would have a larger effect on the aggregate election outcome if it was most effective in counties with more voters. Are the effects of grants larger in populous counties? We evaluate this possibility by splitting our sample into terciles by voting-age population and estimating effects for each subgroup separately. We present our results in Table \ref{tab:vap_tercile} in the online appendix. We find that the effects on turnout and Democratic vote share are no larger in more and less populous counties.

\subsection{Effects Not Larger for Counties Receiving Larger Grants}\label{sec:grant_size}

If additional spending on local election administration increases turnout or Democratic vote share, this would most likely happen because local officials use the money to make it easier for citizens to participate. This implies that the effects of money should increase as they get larger or, at the very least, be unrelated to grant size. While CTCL reports using a formula to decide the maximum amount each county was eligible to receive, the amounts that CTCL distributed to counties ranged from \$0.63 per voting-age resident at the 25th percentile to \$1.38 per voting-age resident at the 75th percentile. Might our small effect estimates mask an effect in counties that receive larger grants?

To answer this question, we split treated counties into three groups based on the amount of grant money going to the county per voting-age resident and produce separate synthetic difference-in-differences estimates of the effect of small, medium, and large grants. We present the results of these analyses in Table \ref{tab:tercile} in the online appendix. Contrary to the expectation that the effect may be limited to places receiving the largest grants, we find that small grants increased Democratic vote share and turnout by more than large grants. We estimate that small grants increased Democratic vote share by 0.62 percentage points and turnout by 0.32 percentage points while large grants decreased Democratic vote share by 0.61 percentage points and turnout by 0.27 percentage points. Given that, unlike our other analyses, the synthetic difference-in-differences weights do a poor job of matching pre-2020 treated and control trajectories for these subgroups, we also present estimates using entropy balancing to match pre-2020 outcome means for control counties to the treated county means. Using entropy balancing, we no longer find negative effects for the largest recipients---the effects are almost exactly zero. We also find slightly smaller effects in places that received smaller grants. These effects average out to approximately the average effect estimates we present in sections \ref{sec:effect} and \ref{sec:alt_est}.

\section{Characterizing the Magnitude of the Effects}\label{sec:magnitude}

How large are the effects of private election administration funding? In this section, we benchmark the magnitude of our effect estimates against the remarkably tight margin of the 2020 presidential election. The 2020 presidential election turned on four states decided by margins of 1.16 percentage points or less: Georgia, Arizona, Wisconsin, and Pennsylvania. The margins in these states were widely understood to be very tight. Are the effects of the private election administration grants as small or smaller than these margins?

One simple way to interpret the effect size is to compare the effect of private election funding on Democratic vote share to the margin in these four close states. Of our three main estimates reported in Table \ref{tab:main}, two of them are too small to have changed the outcome in any state, including Georgia, Arizona, Wisconsin, and Pennsylvania. Our third estimate is about as large as the margin in Wisconsin but still smaller than the margin in Pennsylvania. This is not to say that grant funding was sufficient to change the outcome in any of these states---only a subset of counties received the money, so county-level effects that are roughly the same magnitude as the margin in the state are not large enough to have swung the statewide outcome.

How large are the effects on turnout? One way to interpret these effects is to compare them to the effects of other election administration changes. The effect of the grants on turnout was less than half of the effect of an extra day of early voting \citep{kaplan2020early}, roughly half of the effect of a mailer encouraging citizens to vote by mail \citep{hopkins2021results}, less than one-tenth of the effect of universal vote by mail \citep{gerber2013identifying,thompson2020universal}, and less than one-twentieth of the effect of mobile voting \citep{fowler2020promises}. While the estimated effects of polling place locations on turnout are typically noised than our estimates of the effects of grants, our estimates of the effects of grants tend to be smaller than the effect of having your polling place moved further away \citep{clinton2020polling,tomkins2023blocks,yoder2018polling}. Our estimates of the effects of grants on turnout are small compared to all of these administrative changes.

Given the strong tendency of the grants to go to Democratic-leaning counties, the grants could advantage Democrats more than is implied by the Democratic vote share effect alone. On the other hand, many counties did not receive grants, so the effect of the grants on statewide totals is substantially smaller than the effect in the average county. To account for these concerns, we conducted a simple simulation study. In our simulation, we remove the average effect on turnout and Democratic vote share from all of the treated counties and assume untreated counties remain unchanged.\footnote{We also include municipal-level data from Wisconsin in our simulation analysis.} In states where we do not have grant data, we impute the probability that a county received a grant based on 2016 Democratic vote share and sample 1,000 random possible treatment assignments in those states. We then handle the counties we randomly assigned to treatment in that simulation as we handle the truly treated counties, removing the average effect of the grants on turnout and vote share.

Based on this simple simulation, we find that the estimates from two of our three of our weighted difference-in-differences estimation strategies imply that the grants were too small to swing the outcome of any statewide election. The estimates from our third weighted difference-in-differences strategy are large enough to have changed the outcome of the election in Georgia and Arizona but inconsistent with changing the outcome of the election in Pennsylvania or Wisconsin. Put together, this suggests that, even compared to the margin in very close elections recent elections, the effects of the grants were quite small.

It is important to note that, while these simulations help us understand the magnitude of the effects, we do not intend them as a reflection of what would have happened in the 2020 election had CTCL not made any grants. Our simulations do not account for the general equilibrium effects of the grants, such as changes in partisan spending, get-out-the-vote operations, or other government spending on election administration. Instead, we view our simulations as consistent with our interpretation that the grants had minimal effects.

\section{Discussion}

The large influx of private funding for election administration in 2020, and the fact that Democratic counties were more likely to receive it, has led many politicians, journalists, and pundits to speculate that the funding advantaged Democrats. Despite these widespread concerns, we present evidence that these grants did not substantially increase turnout or Democratic vote share. Our results answer one of the key questions at the center of the debate over private funding of election administration, suggesting that it does not always clearly and substantially favor one party.

Still, our findings leave unanswered two important questions: First, while we find that private funding did not increase turnout, it may have improved the election on other important dimensions. Many of the local officials who received the money said that they would have had trouble reporting their election results on time without the grants. Others said that the money allowed them to hire more staff which may have made the election run more smoothly, made voting more convenient, or improved election security and the accuracy of the count. We cannot observe these effects of the money, but they are important when deciding whether grant programs like these are effective. Second, the large backlash suggests that the grants may have led some citizens to doubt the outcome of the election. If that is the case, it is a potential cost worth considering in future attempts to shore up local election funding.

While our results suggest these grants did not substantially alter the outcome of the 2020 presidential elections, our results do not imply that a grant program like this cannot change election outcomes in the future. We understand our results to be only one input into a broader policy discussion about the appropriateness of grant programs like this given the potential positive effects grants could have on voter experience and security, the potentially harmful effects grants could have on citizen trust in elections, and the risk that future funders could alter election outcomes.

\clearpage
\clearpage

\singlespacing
\bibliographystyle{apsr_fs}
\bibliography{elec_funding.bib}

\clearpage

\doublespacing

\setcounter{section}{0}
\setcounter{table}{0}
\setcounter{figure}{0}
\renewcommand{\thesection}{A.\arabic{section}}
\renewcommand{\thefigure}{A.\arabic{figure}}
\renewcommand{\thetable}{A.\arabic{table}}

\section*{Online Appendix}

Intended for online publication only.

\etocdepthtag.toc{mtappendix}
\etocsettagdepth{mtchapter}{none}
\etocsettagdepth{mtappendix}{subsection}
\tableofcontents

\pagebreak

\section{New Laws Limiting Private Election Administration Grants}

\begin{figure}[h]
\centering
\caption{\textbf{Twenty-Four States Have Passed Laws Limiting Private Election Administration Grants Since 2020.}}\label{fig:bans}
\includegraphics[width=0.8\textwidth]{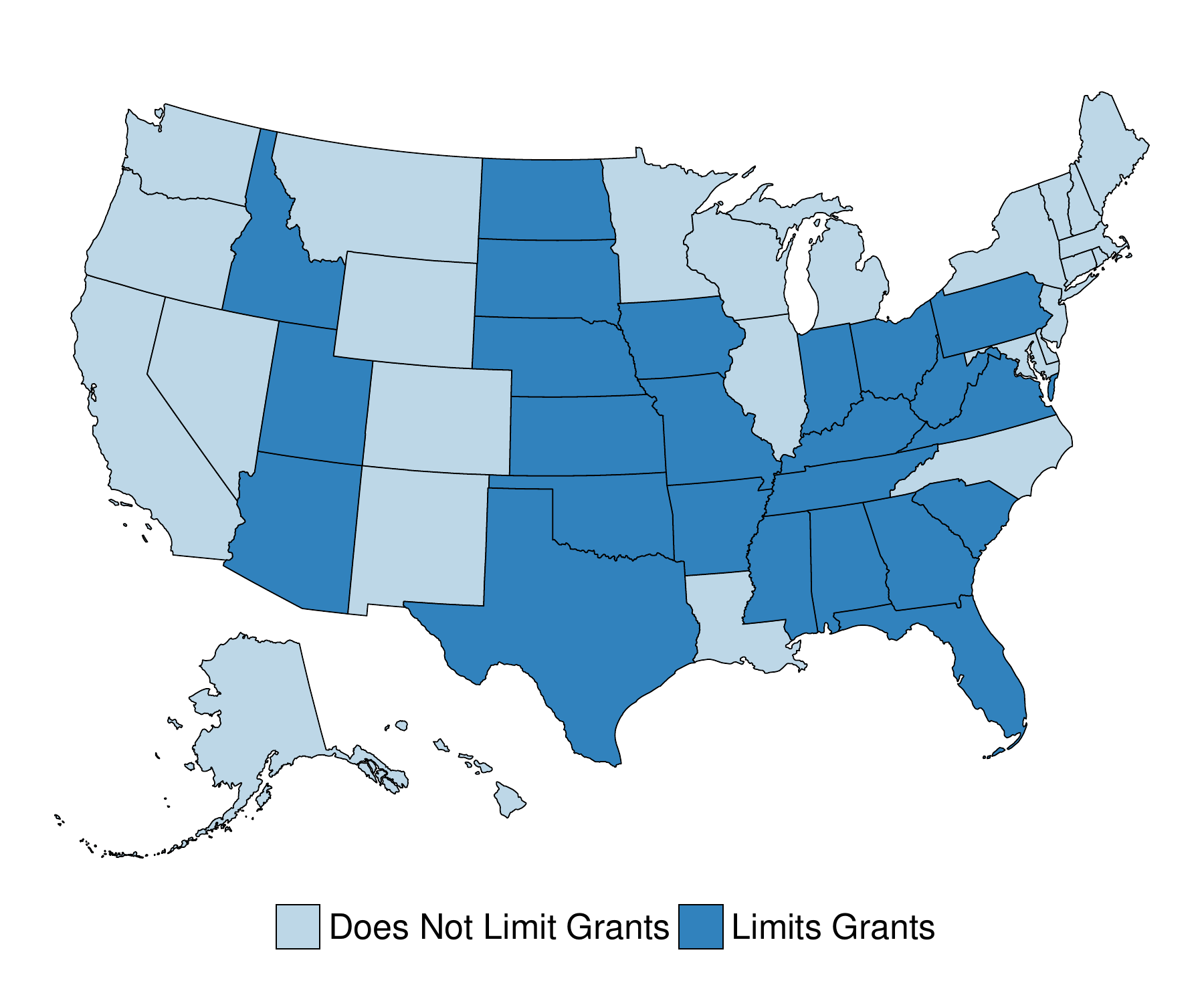}
\end{figure}

As we discuss in Section \ref{sec:intro}, twenty-four states have passed laws banning or substantially limiting private donations that support local election administration. Figure \ref{fig:bans} maps these states. Nearly every Southern state has passed a ban or limit on private funding. The only two Southern states that have not passed such a limit, Louisiana and North Carolina, have Democratic governors who vetoed legislative bills that introduced limits.

\clearpage

\section{Decomposing Grant Selection}\label{sec:selection_covid}

In Section \ref{sec:selection} in the main text, we compare the types of counties that applied for funding to counties that did not apply. In this section, we explore the connection between mail balloting and applying for grant funding.

\begin{figure}[h]
\centering
\caption{\textbf{COVID Cases Primarily Concentrated in Clinton Counties Prior to 2020 Election.}}\label{fig:covid_partisan}
\includegraphics[width=0.8\textwidth]{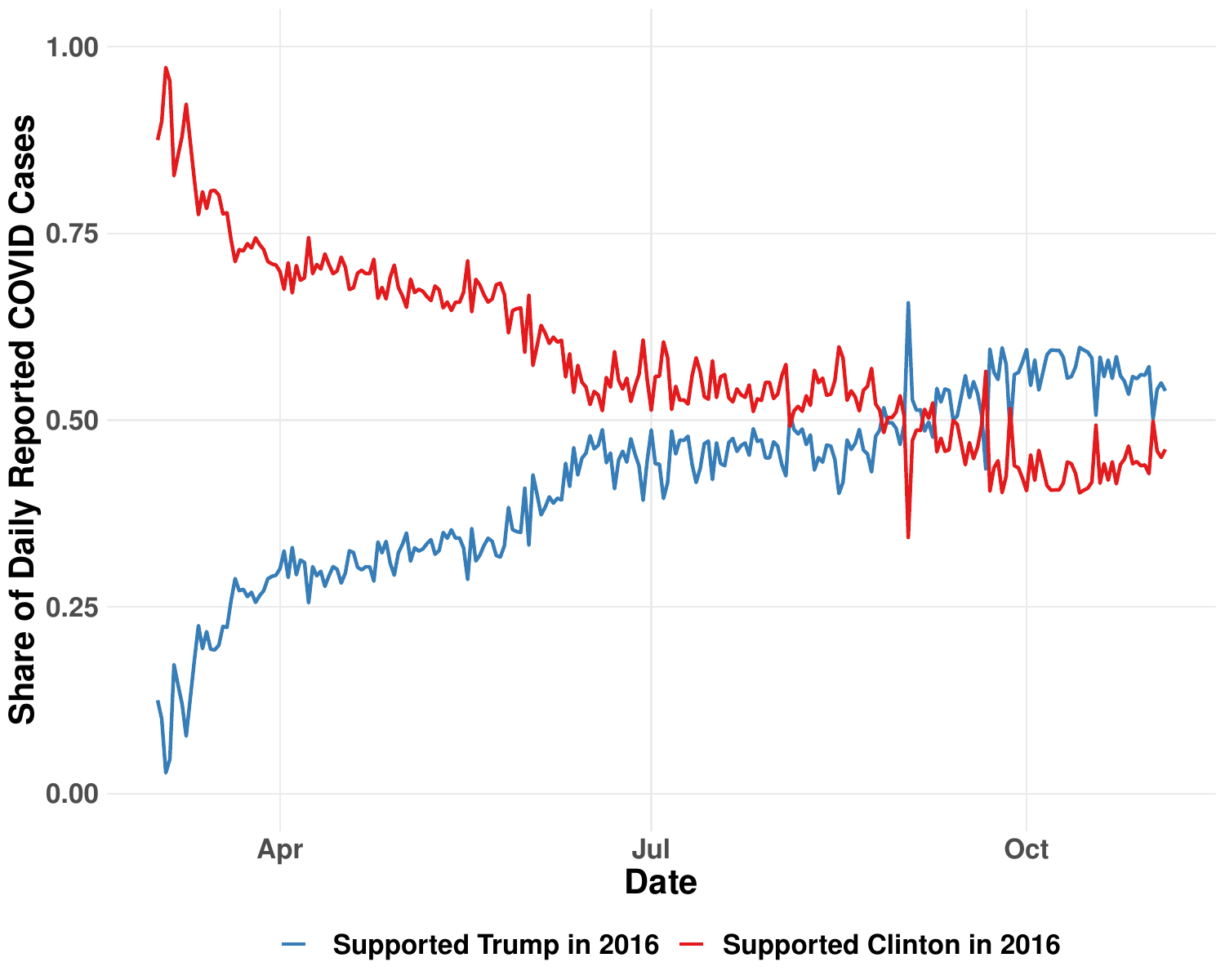}
\end{figure}

During the spring and summer of 2020, COVID killed many more people in Democratic-leaning counties than in Republican-leaning counties. Figure \ref{fig:covid_partisan} captures this pattern. Possibly for this reason as well as other psychological, sociological, political, and economic reasons, Democrats in the public were more eager to vote by mail in the 2020 election than were Republicans. Could this Democratic-Republican gap in demand for mail balloting in 2020 lead to different costs in Democratic and Republican counties and ultimately local officials to expect new costs and apply for a grant to cover them?

\input{figtab/covid_selection_table.tex}

In columns 1 through 3 of Table \ref{tab:selection_covid}, we present suggestive evidence that more citizens in Democratic-leaning counties were eager to vote by mail. We estimate the share of county registrants who intend to vote by mail using UCLA Nationscape survey questions about vote mode linked to the respondent's county. We find that counties that voted more Clinton in 2016 had substantially more residents saying they intended to vote by mail. This relationship shrinks considerably when we include state fixed effects that account different state election laws and is no longer statistically distinguishable from zero at conventional levels, but the point estimate is still positive.

In columns 4 through 6 of Table \ref{tab:selection_covid}, we present suggestive evidence that election officials in Democratic counties were more likely to apply for grants in part because they expected increased costs associated with expanding mail voting. We define three groups of states: those that only allow people to vote by mail with a special excuse, those that send every citizen a mail ballot, and those that removed the need for an excuse to vote by mail in 2020. If officials in Democratic counties were more likely to apply for a grant in part because they anticipated increased mail voting costs, this relationship between Democratic vote share and applying for a grant should be larger in states that expanded mail voting than in states that maintained substantial restrictions on it or already mailed all registered voters a ballot. In column 4 we document that the relationship between Democratic vote share and grant receipt is substantially larger in states that expanded mail voting. In columns 5 and 6, we adjust for additional features of the counties and states. With these adjustments, the relationship between lagged Democratic vote share and grant receipt is more similar in states that expanded mail voting and those that did not, and the difference is no longer statistically distinguishable from zero at conventional levels,  but we the point estimate is still positive and substantively large.

\clearpage

\section{Validating Weighted Diff-in-Diff Using 2016 as a Placebo Treatment Period}\label{sec:placebo}

\input{figtab/placebo_2016_effects_table.tex}

\clearpage

\section{Alternative Strategies for Estimating the Effect of Grants}

\input{figtab/dem_effect_alt_est_table.tex}
\input{figtab/turnout_effect_alt_est_table.tex}

\clearpage

\section{Estimating the Effect of Grants in Wisconsin}\label{sec:wisc_appendix}

We supplement our main county-level analysis with a municipality-level analysis in Wisconsin. We built our main municipality-level analysis dataset from four sources: Wisconsin's Legislative Technology Services Bureau provided election results from 1990 to 2020 at the 2020 municipal ward level. We aggregate this data to the municipal level and link it to a list of all Wisconsin municipalities from Wisconsin's Department of Administration. We add grant amounts by hand to the list of municipalities with geocodes. Finally, we join this data with estimates of the voting age population in each municipality by mapping 2000 and 2010 Census block population statistics into 2020 Census blocks and aggregating to the municipal level. Given that we do not have visibility into how the Legislative Technology Services Bureau computed ward-level election results, we also replicate their work, collecting the original ward-level election results from 2004 to 2020. Our two election datasets are highly correlated and, in many cases, show the exact same number of votes for each party in the same municipality and year.

As we discuss in \ref{sec:wisc}, Table \ref{tab:wi} captures our finding that the grants did not substantially increase Democratic vote share and, if anything, reduced turnout.

\input{figtab/wi_effects_table.tex}

\clearpage

\section{Investigating How Grant Effects Vary Across Counties}

\subsection{Effect of Grants in Battleground States}

\input{figtab/battleground_effects_table.tex}

\clearpage

\subsection{Effect of Grants by County Population Tercile}

\input{figtab/vap_effects_table.tex}

\clearpage

\subsection{Effect of Grants by Grant Size Tercile}

\input{figtab/expterc_effects_table.tex}

\pagebreak

\end{document}

%% file: boilerplate.tex
\usepackage{alltt}
\usepackage[margin=1in]{geometry}
\usepackage{setspace}
\usepackage{color}
\usepackage{graphicx}
\usepackage{multirow}
\usepackage[cm]{fullpage}
\usepackage{enumerate}
\usepackage[fleqn]{amsmath}
\usepackage[all]{xy}
\usepackage{tikz}
\usetikzlibrary{arrows,shapes,backgrounds,decorations.markings}
\usepackage{url}
\usepackage{amssymb}
\usepackage{amsmath}
\usepackage[hang,flushmargin, bottom]{footmisc}
\usepackage[utf8]{inputenc}
\usepackage{sverb}
\usepackage{booktabs}
\usepackage{natbib}
\usepackage{subcaption}
\usepackage{etoc} %
\bibpunct[:~]{(}{)}{;}{a}{}{,}
\usepackage{hyperref}
\usepackage{placeins}

\usepackage{lscape}

\usepackage{colortbl}
\definecolor{Gray}{gray}{0.93}

\usepackage{multirow}

\tikzstyle{block} = [rectangle, draw, 
    text width=5em, text centered, rounded corners, minimum height=4em]
\tikzstyle{line} = [draw, -latex']

\usetikzlibrary{shapes,decorations,arrows,calc,arrows.meta,fit,positioning}
\tikzset{
    -Latex,auto,node distance =1 cm and 1 cm,semithick,
    state/.style ={ellipse, draw, minimum width = 0.7 cm},
    point/.style = {circle, draw, inner sep=0.04cm,fill,node contents={}},
    bidirected/.style={Latex-Latex,dashed},
    el/.style = {inner sep=2pt, align=left, sloped}
}

\usepackage{multibib}
\newcites{App}{Appendix References}

\makeatletter
\renewcommand{\l@section}{\@dottedtocline{1}{1.5em}{2.6em}}
\renewcommand{\l@subsection}{\@dottedtocline{2}{4.0em}{3.6em}}
\makeatother

%% file: figtab/selection_table.tex
\begin{table}[ht]
\centering
\caption{\textbf{Unpacking the Relationship Between Democratic Vote Share and Grant Application.}
\label{tab:selection}}
\begin{tabular}{lcccccc}
\toprule \toprule
 & \multicolumn{6}{c}{Applied for and Received Grant} \\
 & (1) & (2) & (3) & (4) & (5) & (6) \\
\midrule
Lag Dem Vote Share & 0.69 & 0.55 & 0.31 & 0.44 & 0.42 & 0.42 \\
 & (0.06) & (0.07) & (0.07) & (0.10) & (0.10) & (0.10) \\[2mm]
Log(Population) & & & 0.06 & 0.06 & 0.06 & 0.06 \\
 & & & (0.01) & (0.01) & (0.01) & (0.01) \\[2mm]
Log(Median Income) & & & 0.10 & 0.06 & 0.06 & 0.06 \\
 & & & (0.05) & (0.05) & (0.05) & (0.05) \\[2mm]
Metro & & & & 0.03 & 0.03 & 0.03 \\
 & & & & (0.02) & (0.02) & (0.02) \\[2mm]
Non-Hisp White Share & & & & 0.16 & 0.13 & 0.13 \\
 & & & & (0.09) & (0.09) & (0.09) \\[2mm]
COVID Death Rate & & & & & -0.00 & -0.00 \\
 & & & & & (0.02) & (0.02) \\[2mm]
Social Distancing Share & & & & & -0.02 & -0.02 \\
 & & & & & (0.03) & (0.03) \\[2mm]
NACo & & & & & & 0.01 \\
 & & & & & & (0.02) \\[2mm]
Constant & 0.14 & & & & & \\
 & (0.02) & & & & & \\[2mm]
Observations & 2594 & 2594 & 2593 & 2592 & 2537 & 2537 \\
\midrule
State FEs & No & Yes & Yes & Yes & Yes & Yes \\
\bottomrule \bottomrule
\multicolumn{7}{p{0.8\textwidth}}{\footnotesize Robust standard reported in parentheses. 
Population is the voting-age population. 
Median Income is median household income measured with the 5-year ACS ending in 2019. 
Metro is an indicator for urban and suburban counties based on the Census nine-value urban-rural continuum. 
Non-Hisp White Share is the share of residents who are classified as non-Hispanic white in the 2020 census. 
Covid death rate is the number of deaths per 1,000 residents prior to September 1, 2020. 
Social Distancing Share is the share of Nationscape respondents in the county who report always complying 
with recommended social distancing in the early fall of 2020. 
NACo is an indicator for county membership in the National Association of Counties. 
}
\end{tabular}
\end{table}

%% file: figtab/main_effects_table.tex
\begin{table}[t]
\centering
\caption{\textbf{Election Administration Grants Did Not Noticeably Advantage Democrats or Increase Turnout in 2020.}
\label{tab:main}}
\begin{tabular}{lcccc|cccc}
\toprule \toprule
 & \multicolumn{4}{c}{Dem Vote Share (\%)} & \multicolumn{4}{c}{Turnout (\%)} \\
 & (1) & (2) & (3) & (4) & (5) & (6) & (7) & (8) \\
\midrule
Grant Recipient in 2020 &  3.24 &  0.36 &  0.10 &  0.02 &  0.24 & -0.03 &  0.13 &  0.03 \\
 & (0.34) & (0.11) & (0.12) & (0.11) & (0.21) & (0.15) & (0.14) & (0.14) \\[2mm]
Num Grant Recipients  & 924 & 924 & 924 & 924 & 924 & 924 & 924 & 924 \\
Num Counties  & 2,597 & 2,597 & 2,597 & 2,597 & 2,597 & 2,597 & 2,597 & 2,597 \\
Observations  & 20,776 & 20,776 & 20,776 & 20,776 & 20,776 & 20,776 & 20,776 & 20,776 \\
\midrule
County FEs & Yes & Yes & Yes & Yes & Yes & Yes & Yes & Yes \\
Year FEs & Yes & Yes & Yes & Yes & Yes & Yes & Yes & Yes \\
County Weights & No & No & Yes & Yes & No & No & Yes & Yes \\
Year Weights & No & Yes & No & Yes & No & Yes & No & Yes \\
\bottomrule \bottomrule
\multicolumn{9}{p{1\textwidth}}{\footnotesize Standard errors estimated with 1,000 county block 
bootstrap samples reported in parentheses. Data is a balanced panel of counties in the 8 presidential 
elections from 1992 to 2020.}
\end{tabular}
\end{table}

%% file: figtab/covid_selection_table.tex
\begin{table}[ht]
\centering
\caption{\textbf{In Democratic-Leaning Counties, More People Intended to Vote-by-Mail and Officials Were More Likely to Apply for CTCL Grant.}
\label{tab:selection_covid}}
\begin{tabular}{lccc|ccc}
\toprule \toprule
 & \multicolumn{3}{c}{Mail Voting Share} & \multicolumn{3}{c}{Received Grant} \\
 & (1) & (2) & (3) & (4) & (5) & (6) \\
\midrule
Lag Dem Vote Share & 0.23 & 0.33 & 0.09 & 0.43 & 0.15 & 0.44 \\
 & (0.04) & (0.06) & (0.07) & (0.11) & (0.13) & (0.15) \\[2mm]
Lag Dem Vote Share $\times$ & & & & 0.54 & 0.35 & 0.26 \\
 \ \ Mail Voting Expansion & & & & (0.18) & (0.18) & (0.17) \\[2mm]
Mail Voting Expansion & & & & -0.10 & -0.01 & \\
 & & & & (0.06) & (0.07) &  \\[2mm]
Observations & 2356 & 2355 & 2355 & 1238 & 1237 & 1237 \\
\midrule
Controls & No & Yes & Yes & No & Yes & Yes \\
State FEs & No & No & Yes & No & No & Yes \\
\bottomrule \bottomrule
\multicolumn{7}{p{0.8\textwidth}}{\footnotesize Robust standard reported in parentheses. 
Mail voting share is the weighted share pre-election NationScape respondents in a county who 
said they intended to vote by mail. No-excuse expansion codes states that enacted no-excuse 
absentee voting for the first time in 2020 as 1. States that already had universal mail ballot 
delivery or offered absentee voting only with an excuse in 2020 are coded as 0. All other states 
are held out of this analysis. The controls are the log of voting age population, the log of 
median household income, and the non-Hispanic white share of the population.}
\end{tabular}
\end{table}

%% file: figtab/placebo_2016_effects_table.tex
\begin{table}[ht]
\centering
\caption{\textbf{Weighted Difference-in-Differences Approach Balances Dem Vote Share and Turnout in 2016.}
\label{tab:placebo2016}}
\begin{tabular}{lcccc|cccc}
\toprule \toprule
 & \multicolumn{4}{c}{Dem Vote Share (\%)} & \multicolumn{4}{c}{Turnout (\%)} \\
 & (1) & (2) & (3) & (4) & (5) & (6) & (7) & (8) \\
\midrule
Grant Recipient in 2016 &  3.38 &  0.81 &  0.48 &  0.38 &  0.12 & -0.14 & -0.20 & -0.17 \\
 & (0.36) & (0.22) & (0.23) & (0.23) & (0.17) & (0.10) & (0.10) & (0.10) \\[2mm]
Num Grant Recipients  & 924 & 924 & 924 & 924 & 924 & 924 & 924 & 924 \\
Num Counties  & 2,594 & 2,594 & 2,594 & 2,594 & 2,594 & 2,594 & 2,594 & 2,594 \\
Observations  & 18,158 & 18,158 & 18,158 & 18,158 & 18,158 & 18,158 & 18,158 & 18,158 \\
\midrule
County FEs & Yes & Yes & Yes & Yes & Yes & Yes & Yes & Yes \\
Year FEs & Yes & Yes & Yes & Yes & Yes & Yes & Yes & Yes \\
County Weights & No & No & Yes & Yes & No & No & Yes & Yes \\
Year Weights & No & Yes & No & Yes & No & Yes & No & Yes \\
\bottomrule \bottomrule
\multicolumn{9}{p{1\textwidth}}{\footnotesize Standard errors estimated with 1,000 county block 
bootstrap samples reported in parentheses. Data is a balanced panel of counties in the 7 presidential 
elections from 1992 to 2016.}
\end{tabular}
\end{table}

%% file: figtab/dem_effect_alt_est_table.tex
\begin{table}[ht]
\centering
\caption{\textbf{Election Administration Grants Did Not Noticeably Advantage Democrats in 2020, Alternative Estimators.}
\label{tab:alt_est_dem}}
\begin{tabular}{lccccc}
\toprule \toprule
 & \multicolumn{5}{c}{Dem Vote Share (\%)} \\
 & (1) & (2) & (3) & (4) & (5) \\
\midrule
Grant Recipient in 2020 & 0.28 & 0.40 & 0.23 & 0.25 & 0.15 \\
 & (0.13) & (0.77) & (0.11) & (0.11) & (0.07) \\[2mm]
Num Grant Recipients  & 924 & 924 & 924 & 924 & 924 \\
Num Counties  & 2,597 & 2,597 & 2,597 & 2,597 & 2,597 \\
Observations  & 20,776 & 20,776 &  2,597 &  2,597 &  2,597 \\
\midrule
Estimator & SDID Without & Synthetic & Entropy & Causal & Super \\
 & Intercept & Control & Balancing & Forest & Learner \\
\bottomrule \bottomrule
\multicolumn{6}{p{0.9\textwidth}}{\footnotesize Standard errors estimated with 1,000 county block 
bootstrap samples reported in parentheses. Data for columns 1, 2, 6, and 7 is a balanced panel of counties in the 8 presidential 
elections from 1992 to 2020. Data for columns 3, 4, 5, 8, 9, and 10 is wide with 7 lags of the dependent variable included. 
Synthetic control is a regularized synthetic control. SDID without intercept is synthetic difference-in-differences without county fixed effects. 
Entropy balancing is maximum entropy reweighting to balance grant recipients and non-recipients on the average of each lag of the outcome. 
Causal forest is double machine learning using random forests for both the outcome and propensity models. 
Super learner is double machine learning using an ensemble learner for both the outcome and propensity models.}
\end{tabular}
\end{table}

%% file: figtab/turnout_effect_alt_est_table.tex
\begin{table}[ht]
\centering
\caption{\textbf{Election Administration Grants Did Not Noticeably Increase Turnout in 2020, Alternative Estimators.}
\label{tab:alt_est_turnout}}
\begin{tabular}{lccccc}
\toprule \toprule
 & \multicolumn{5}{c}{Turnout (\%)} \\
 & (1) & (2) & (3) & (4) & (5) \\
\midrule
Grant Recipient in 2020 &  0.03 & -0.02 &  0.47 &  0.00 &  0.04 \\
 & (0.14) & (0.39) & (0.39) & (0.14) & (0.12) \\[2mm]
Num Grant Recipients  & 924 & 924 & 924 & 924 & 924 \\
Num Counties  & 2,597 & 2,597 & 2,597 & 2,597 & 2,597 \\
Observations  & 20,776 & 20,776 &  2,597 &  2,597 &  2,597 \\
\midrule
Estimator & SDID Without & Synthetic & Entropy & Causal & Super \\
 & Intercept & Control & Balancing & Forest & Learner \\
\bottomrule \bottomrule
\multicolumn{6}{p{0.9\textwidth}}{\footnotesize Standard errors estimated with 1,000 county block 
bootstrap samples reported in parentheses. Data for columns 1, 2, 6, and 7 is a balanced panel of counties in the 8 presidential 
elections from 1992 to 2020. Data for columns 3, 4, 5, 8, 9, and 10 is wide with 7 lags of the dependent variable included. 
Synthetic control is a regularized synthetic control. SDID without intercept is synthetic difference-in-differences without county fixed effects. 
Entropy balancing is maximum entropy reweighting to balance grant recipients and non-recipients on the average of each lag of the outcome. 
Causal forest is double machine learning using random forests for both the outcome and propensity models. 
Super learner is double machine learning using an ensemble learner for both the outcome and propensity models.}
\end{tabular}
\end{table}

%% file: figtab/wi_effects_table.tex
\begin{table}[ht]
\centering
\caption{\textbf{Election Administration Grants Did Not Noticeably Advantage Democrats or Increase Turnout in Wisconsin in 2020.}
\label{tab:wi}}
\begin{tabular}{lcccc|cccc}
\toprule \toprule
 & \multicolumn{4}{c}{Dem Vote Share (\%)} & \multicolumn{4}{c}{Turnout (\%)} \\
 & (1) & (2) & (3) & (4) & (5) & (6) & (7) & (8) \\
\midrule
Grant Recipient in 2020 &  1.33 &  0.31 &  0.24 &  0.11 & -0.64 & -0.71 & -0.70 & -0.69 \\
 & (0.58) & (0.20) & (0.19) & (0.19) & (0.40) & (0.28) & (0.28) & (0.29) \\[2mm]
Num Grant Recipients  & 206 & 206 & 206 & 206 & 206 & 206 & 206 & 206 \\
Num Counties  & 1,843 & 1,843 & 1,843 & 1,843 & 1,843 & 1,843 & 1,843 & 1,843 \\
Observations  & 12,901 & 12,901 & 12,901 & 12,901 & 12,901 & 12,901 & 12,901 & 12,901 \\
\midrule
County FEs & Yes & Yes & Yes & Yes & Yes & Yes & Yes & Yes \\
Year FEs & Yes & Yes & Yes & Yes & Yes & Yes & Yes & Yes \\
County Weights & No & No & Yes & Yes & No & No & Yes & Yes \\
Year Weights & No & Yes & No & Yes & No & Yes & No & Yes \\
\bottomrule \bottomrule
\multicolumn{9}{p{1\textwidth}}{\footnotesize Standard errors estimated with 1,000 county block 
bootstrap samples reported in parentheses. Data is a balanced panel of counties in the 8 presidential 
elections from 1992 to 2020.}
\end{tabular}
\end{table}

%% file: figtab/battleground_effects_table.tex
\begin{table}[ht]
\centering
\caption{\textbf{Election Administration Grants Did Not Noticeably Advantage Democrats or Increase Turnout in 2020, Battleground States.}
\label{tab:battleground}}
\begin{tabular}{lcc|cc}
\toprule \toprule
 & \multicolumn{2}{c}{Dem Vote Share (\%)} & \multicolumn{2}{c}{Turnout (\%)} \\
 & Close & Cook & Close & Cook \\
 & States & Battlegrounds & States & Battlegrounds \\
\midrule
Grant Recipient in 2020 &  0.54 &  0.04 &  0.10 & -0.46 \\
 & (0.29) & (0.21) & (0.32) & (0.22) \\[2mm]
Num Grant Recipients  & 119 & 326 & 119 & 326 \\
Num Counties  & 421 & 906 & 421 & 906 \\
Observations  & 3,368 & 7,248 & 3,368 & 7,248 \\
\midrule
County FEs & Yes & Yes & Yes & Yes \\
Year FEs & Yes & Yes & Yes & Yes \\
County Weights & Yes & Yes & Yes & Yes \\
Year Weights & Yes & Yes & Yes & Yes \\
\bottomrule \bottomrule
\multicolumn{5}{p{.8\textwidth}}{\footnotesize Standard errors estimated with 1,000 county block 
bootstrap samples reported in parentheses. Data is a balanced panel of counties in the 8 presidential 
elections from 1992 to 2020. Close states are those where the winner was decided by fewer than 5 percentage points. Cook battleground states are those that the Cook Political Report identified as battlegrounds prior to election day.}
\end{tabular}
\end{table}

%% file: figtab/vap_effects_table.tex
\begin{table}[ht]
\centering
\caption{\textbf{Election Administration Grants Did Not Noticeably Advantage Democrats or Increase Turnout in 2020, Voting Age Population Tercile.}
\label{tab:vap_tercile}}
\begin{tabular}{lccc|ccc}
\toprule \toprule
 & \multicolumn{3}{c}{Dem Vote Share (\%)} & \multicolumn{3}{c}{Turnout (\%)} \\
 & Small & Medium & Large & Small & Medium & Large \\
 & Population & Population & Population & Population & Population & Population \\
\midrule
Grant Recipient in 2020 & -0.24 &  0.15 & -0.42 & -0.71 &  0.08 & -0.14 \\
 & (0.17) & (0.13) & (0.21) & (0.24) & (0.23) & (0.23) \\[2mm]
Num Grant Recipients  & 256 & 253 & 415 & 256 & 253 & 415 \\
Num Counties  & 866 & 865 & 865 & 866 & 865 & 865 \\
Observations  & 6,928 & 6,920 & 6,920 & 6,928 & 6,920 & 6,920 \\
\midrule
County FEs & Yes & Yes & Yes & Yes & Yes & Yes \\
Year FEs & Yes & Yes & Yes & Yes & Yes & Yes \\
County Weights & Yes & Yes & Yes & Yes & Yes & Yes \\
Year Weights & Yes & Yes & Yes & Yes & Yes & Yes \\
\bottomrule \bottomrule
\multicolumn{7}{p{1.1\textwidth}}{\footnotesize Standard errors estimated with 1,000 county block 
bootstrap samples reported in parentheses. Data is a balanced panel of counties in the 8 presidential 
elections from 1992 to 2020. Small, medium, and large population counties are defined by terciles of 
voting age population.}
\end{tabular}
\end{table}

%% file: figtab/expterc_effects_table.tex
\begin{table}[ht]
\centering
\caption{\textbf{Election Administration Grants Did Not Noticeably Advantage Democrats or Increase Turnout in 2020, Grant Size Tercile.}
\label{tab:tercile}}
\begin{tabular}{lccc|ccc}
\toprule \toprule
 & \multicolumn{3}{c}{Dem Vote Share (\%)} & \multicolumn{3}{c}{Turnout (\%)} \\
 & Small & Medium & Large & Small & Medium & Large \\
 & Grant & Grant & Grant & Grant & Grant & Grant \\
\midrule
Grant Recipient in 2020 &  0.62 &  0.08 & -0.61 &  0.32 &  0.07 & -0.27 \\
 & (0.12) & (0.14) & (0.18) & (0.21) & (0.21) & (0.22) \\[2mm]
Num Grant Recipients  & 305 & 304 & 315 & 305 & 304 & 315 \\
Num Counties  & 1,978 & 1,977 & 1,988 & 1,978 & 1,977 & 1,988 \\
Observations  & 15,824 & 15,816 & 15,904 & 15,824 & 15,816 & 15,904 \\
\midrule
County FEs & Yes & Yes & Yes & Yes & Yes & Yes \\
Year FEs & Yes & Yes & Yes & Yes & Yes & Yes \\
County Weights & Yes & Yes & Yes & Yes & Yes & Yes \\
Year Weights & Yes & Yes & Yes & Yes & Yes & Yes \\
\bottomrule \bottomrule
\multicolumn{7}{p{0.9\textwidth}}{\footnotesize Standard errors estimated with 1,000 county block 
bootstrap samples reported in parentheses. Data is a balanced panel of counties in the 8 presidential 
elections from 1992 to 2020. Grant sizes are determined by tercile of grant size per voting age resident among recipients. Small is the smallest tercile, and large is the largest tercile.}
\end{tabular}
\end{table}